**MICHELE TUCCI**

**DEPARTMENT OF PUBLIC ECONOMICS**
**FACULTY OF ECONOMICS**
**UNIVERSITY OF ROME "LA SAPIENZA"**

# EVOLUTION AND GRAVITATION: A COMPUTER SIMULATION OF A NON-WALRASIAN EQUILIBRIUM MODEL

*Economic dynamics "in the mist over troubled waters"*

**Rome – 2002**

**Abstract**
The following notes contain a computer simulation concerning a basic non-Walrasian equilibrium system, following the Edmond Malinvaud "short side" approach, as far as the price adjustment is concerned, and the sequential Hicksian "weeks" structure with regard of the temporal characterization. Two strongly heterogeneous classes of agents will be taken into consideration: the "rich" and the "poor" and it will be shown how such sociological distinction critically influences the evolution of the framework under consideration. Finally, some methodological consideration will be added, regarding the comparison between the evolutionary and the gravitational paradigms and the evaluation of approaches belonging to rival schools of economic thought.

# EVOLUTION AND GRAVITATION: A COMPUTER SIMULATION OF A NON-WALRASIAN EQUILIBRIUM MODEL

## *Economic dynamics "in the mist over troubled waters"*

## **Michele Tucci**[1]

## 1. Introduction

The following notes are to be considered as the synthesis of some results pertaining to a research project which has been carried out by the author for quite a long time. The main point of such an activity has been focused on the formal reconstruction of the approaches put forward by the major schools of economic thought and, above all, on the possibility to build "translations" linking such alternative interpretations of economic phenomena[2]. As far as the operational side of the matter is concerned, the interest has been concentrated on defining the evolutionary paradigm of economic thought, as opposed to the gravitational one, and on building feasible tools to cope with such a realm of reality[3]. The second class of issues is to be dealt with in the following paragraphs. The argumentation will be kept as informal as possible, since the scope is to give a general view of the matter, rather than to dig into details, which have been treated by the author in more technical papers. Still, in the third paragraph a simple computer simulation will be shown.

Let's start these notes with a dilemma. Not of theoretical nature, but an operative one: let's examine the general framework underlying the process of decision making in the field of economic events. Firstly, of course, we suppose to have clearly defined the target we want to reach. In order to pursue such a scope, let's suppose that we are able to access a given information set and to use computational tools which can be supposed up to the state of art. Clearly, it is possible that, though operating at the best of our analytical capabilities, we are unable to define a unique strategy, but on the contrary we come up with a multiplicity of potentially optimal solutions, each one conditionally subdued to the future outcome of a number of critical events. In other words, each potential strategy may be optimal, or may be not, according to the fact that certain conditions will, or will not, be verified in the future within the economic environment where we have to operate. Since we are neither gods nor magicians, and therefore the future is obscure for us as a black cat in a dark night, we are left with a major problem or, to be more precise, a dilemma. Dealing with it in general terms, we face three possibilities. The first one is to do nothing. In fact, we could decide that if we were unable to define a strategy which would be successful under any condition, i.e. for any possible outcome of the future critical events, it is better not to act. Even if such a choice cannot be considered an irrational one, often it is self-defeating, since it put the operator in a completely passive stance: the world will be taken "as it is", without any possibility to influence the course of the events.

[1] Michele Tucci, Dipartimento di Economia Pubblica – Facoltà di Economia – Università degli Studi di Roma "La Sapienza", v. del Castro Laurenziano 9, 00161 Roma, Italy. Email: tucci@dep.eco.uniroma1.it
[2] See Tucci (1977), (1988a), (1992), (1997); Pearce-Tucci (1982).

The second approach is the one that we will define of a gravitational type. Since we may reasonably suppose that the past and the present can always be known if we make the effort to collect the desired records, what we can do is to extend our initially given information set, in such a way to include any interesting item of knowledge concerning the past and the present events of the economic environment under examination. Then, using the best statistical inference tools, we can try to forecast the future values of the parameter related to the above mentioned critical events. Therefore, now we should be able to select the supposed optimal strategy among those belonging to the above defined set. The week point of such an approach lays in the unavoidable gravitational assumption: we must suppose that the structural features of the economic world where we operate do not change in the course of time. Any entrepreneur knows well that such a conjecture is highly abstract. In economic matters the future rarely resembles the past and therefore critical choices can hardly ever be taken only according to historical records.

We are now left with the third choice: the evolutionary approach. Along with Keynes' treatment of the "animal spirits" expectations, the entrepreneur will base his investment acts upon his own subjective views relative to the future of the economic world where he is operating. Creating a subjective model of such a setting is a demanding task which implies both quantitative and qualitative aspects. As far as the quantitative estimates are concerned, we can relay on De Finetti approach based on the definition of subjective probability[4]. In such a way we are able to subjectively measure the outcome of critical events. Still, in order to obtain a structure that can be used to forecast the evolutionary changes of the economic system, we need to model the complex structure of relationships which interconnects each element of the environment to every other one. To accomplish such a task, we can use Simon's simulacra[5]: a computer simulation of the economic environment we are interested in. Once we possess such an instrument, and we have subjectively estimated the parameters which are needed to operate the simulation, we can bring into life our little virtual world. In other words, we can let the application run and observe how the critical variables of the model evolve through time. If the procedure has been carried on well enough, we might be able to obtain a forecast of the evolutionary changes that will take place in the economic context under examination[6]. With regard to such a goal, two points need to be kept in mind. Firstly, we should remember Keynes' stress on "short period". Expectations founded on "animal spirits", as well as subjective probabilistic estimates, are meaningful only on a "short run" basis, while they are useless if we pretend reliability in an asymptotic context. Such observation is supported by Simon's continuity principle which can be resumed as it follows: the more a simulation is close to the portion of reality we are analyzing, the more its evolutionary behavior will be similar to what will happen in the real world. Nevertheless, since no simulation can be identical to the simulated phenomena, the more we move further in the time scale, the more the evolution of the simulation will differ from the actual events. As a consequence of such a principle, limit patterns cannot be analyzed by using Simon's tool. Therefore, asymptotic behavior of a simulation model ought to be regarded as meaningless.

---

[4] See De Finetti (1958), (1964), (1968).
[5] See Simon (1992), (2000); Ando-Fisher-Simon (1963).

Secondly, it should be noted that, even within an environment which shows strong gravitational properties, Simon's simulacra could lead to useful results. In fact, short term properties of gravitational economic systems cannot be deducted from the asymptotic outcome of the model, since the path leading to a limit point is generally influenced by a great variety of contingent events. Therefore, while a limit point, when is endorsed with sufficient stability properties, could constitute the solution of a gravitational problem, the path leading to such a solution will be influenced by the evolutionary conditions which are prevailing locally. In such a contest, the above outlined simulation methodology is probably the only one susceptible to provide "short run" operative predictions.

In the following paragraph some aspects concerning the evolutionary changes of an economic environment will be analyzed, while the third one will include the outcome of a simple computer simulation model. In the forth paragraph it will be debated whether the general equilibrium paradigm may be considered as belonging to the gravitational realm. Finally, the last one will contain some conclusive remarks.

---

[6] For an example of such a procedure see Kirman-Vriend (2000).

## 2. On the nature of evolutionary processes

Let's write down some specifications about what can be defined as an economic evolutionary structure. Let's take into consideration a dynamical process, without worrying about the mathematical or the computational tools which are used to define such an entity. It could be a differential equation or a finite difference one, as well as a more complex topological construction or a C++ source code: the following comments are largely independent from such a choice. Let's suppose that the reader is familiar with the idea and the subtleties of what is defined as "attraction point" and let's proceed to formulate some propositions concerning the same. In general terms, we might agree on the following statements:

*(1). An attraction point is a globally stable solution to a dynamics problem.*

*(2). A lesser form of attraction point is represented by a set of locally stable solutions. In this case, there may be chaotic dynamics. For example, such a instance occurs when the location of a dynamical object falls on the boundary between two areas of local stability. Thus, the slightest disturbance is able to critically influence the outcome of the dynamical process.*

If we take into consideration dynamical contexts which include attraction points as they have been defined in (1) or (2), then we should agree that those cases belong to the gravitational world, even if in the second one it could be extremely difficult, or even impossible, to forecast the actual trajectory of the dynamical object. Chaotic dynamics ought to be considered as an extreme case of gravitational dynamics, while evolutionary processes show a completely different pattern of behavior.

The evolutionary point of view is clearly expressed in Darwin writings and, as far as economics is concerned, in the reconstruction operated by Schumpeter[7]. The main difference between a gravitational and an evolutionary dynamics is that the latter includes some critical events which cannot be deducted from the initially given framework used to define the context under examination. In the classical Darwinian evolution process, based on the binomial condition "appearance of the mutation – selection of the fittest", the unpredictable element is represented by the coming into existence of the mutation, while generally the selection process is suitable of some kind of formal reconstruction[8]. In the field of economics, the Keynesian notion of "animal spirits" leads to still another class of critical events. Whenever an entrepreneur take an investment decision, according to his subjective probabilistic estimations and to his own vision of the world, such an action is not deducible from an *a priori* calculation. Therefore, this class of occurrences constitutes a category of  critical events which is peculiar to the world of economics and represents one of the founding concept of the Keynesian school of thought. In short, the world of economics admits at least two classes of phenomena which cannot be analyzed by the usage of gravitational models: the mutations in the "appearance of the mutation – selection of the fittest" framework, and the "animal spirits" based on a subjective view of the environment where the entrepreneur is due to operate.

---

[7] See Schumpeter (1971), (1977), (1982).

Besides the above listed cases (1) and (2), let's now take into consideration some other forms of the attraction point which will be shown to belong to the evolutionary realm.

*(3). Let's consider case (1). Let's suppose that the attraction point, instead of been given once and for all, changes through time, in a manner that shows some continuity properties. Due to this last assumption, there might be the possibility that the evolutionary changes of the structure could be estimated by the usage of sequences of samplings fit to catch the change in the data generating process. This case could be classified as belonging to the field of "small" evolution,*

---

[8] See Tucci (2000), Tucci-Barchetti (2001).

*since we can still try to use some of the usual gravitational tools in order to obtain at least a first estimate of the actual phenomena.*

*(4). Let's focus our attention on a context similar to the above described case (3), but such that the attraction point mutates through time with indefinite modes, thus preventing the possibility to formulate any a priori assumption concerning the nature of the alteration. This occurrence could be classified as belonging to the realm of "big" evolution, as compared with the "small" one taken into consideration in the former instance. The way to deal with this case implies the usage of De Finetti's subjective probability approach and Simon's simulacra.*

From what has been shown up to here one question rises quite naturally: why cannot we analyze every phenomenon by means of a gravitational framework? Why certain phenomena seem to fit a gravitational reconstruction very easily, while others do not follow the same pattern of behavior? Is it simply a matter of lacking information or there is more to it? The solution of such a problem is a complex one, as it involves some philosophical arguments. In fact, the expression "lack of information" is rather ambiguous, since it may imply the existence of pieces of knowledge which I do not possess, but I may eventually gather in the future; or it could refer to knowledge that can never be obtained, due to its intrinsic nature. The scope of scientific research is to expand the boundary of knowledge, but there is no prove that we can eventually obtain any piece of information we seek. As it has been already pointed out above, within the field of economics there are at least two phenomena about which the hope to obtain gravitational interpretations is rather deem: the formulation of "animal spirits" expectations and the emergence of mutations in a Darwinian evolutionary process. As it is well known, the first issue was analyzed by Keynes, while the second one has been reconstructed by Schumpeter. Both topics have been analyzed in details by the author of the present notes and therefore they will not be treated here[9]. In short, since "animal spirits" expectations consist on a subjective conjecture about the future state of the world cast by each entrepreneur, the possibility to built a gravitational law governing such an expression of human personality is rather feeble. As far as Schumpeter reconstruction is concerned, the evolution of the productive technology is based on two main phases: the emergence of mutations – i.e. the appearance of substantially new techniques – and the selection of the fittest – i.e. the market operated process of individuating the most economically profitable production line. While the latter phase can be formalized, provided that we have the necessary information, the former depends on a mixture of human creativity, favorable conditions and pure luck, up to the point that events of such a nature are nearly impossible to forecast, as well as to artificially induce.

In concluding the present paragraph, let's take into consideration the multitude of economic phenomena which are grouped under the label of economic crises. The analysis of such a topic is strictly related with what has been treated above, since after all a crisis is just a particularly swift economic change, usually with negative implications. Therefore, the general framework which has been sketched still holds. More in details, it is worth to add some considerations on the nature of crises, whether they ought to be regarded as local or global ones.

A local crisis arises within an otherwise stable context and it exists in a limited subset of the initially given environment.  Local crises are linked to such an environment by a set of identifiable status variables in such a way that the inside-outside interaction between the two areas can be dealt in formal terms. Therefore, local crises can be controlled in a reasonably simple way by creating a simulation model – or a traditional econometric model – and finding the suitable policies for the control variables, i.e. the subset of the input-output variable that we choose as operative tools.

---

[9] See Tucci (1988b), (2000).

Global crises are quite a different business. In facts, in such a case we cannot distinguish between internal and external areas, since the global nature of the crisis does not allow us to establish such a distinction. Operating on a single variable is often ineffective – if not dangerous – because of feedback effects. Trying to control a numerous set of variables at once is extremely difficult, due to the size of phenomena. Global crises usually evolve to significant structural changes, and more often than not there is very little capacity of control. From what has been said up to now it should be clear that generally local crises belong to the gravitational realm, while global ones are part of the evolutionary world.

In the following paragraph a simple computer simulation will be shown. It will concern a model based on Hicksian "weeks"[10] and non-Walrasian equilibria, following the treatment due to Malinvaud[11]. Thought such an environment cannot be defined as an evolutionary setting in the full meaning of the world, since it lacks the above mentioned requirements, still it lays a step away from the gravitational world. In fact, "steady states" and asymptotic conditions are completely neglected, while the stress is toward the patterns of behavior put forward by the "short run" simulation dynamics.

---

[10] See Hicks (1959), (1971), (1985).
[11] See Malinvaud (1953), (1977); Arcelli (1980).

**3. The simulation model**

The model we will employ in the present paragraph is a very simple and unsophisticated non-Walrasian equilibrium system, following the Edmond Malinvaud approach, as far as the price adjustment is concerned, and the sequential Hicksian "weeks" structure with regard of the temporal characterization[12]. Only three commodities will be taken into consideration: a consumer good, a capital good and "free time", i.e. time that is subtracted from the working schedule for the sake of engaging different types of activities. Two single-production lines will be operated, turning out consumer and capital goods. Production employs as inputs labor and capital good; for the sake of simplicity, capital will be considered as a circulating commodity, i.e. it will be destroyed throughout a single production cycle, while both utility and production functions will be represented in a constant returns to scale Cobb-Douglas form. Two strongly heterogeneous classes of agents will be taken into consideration: the "rich" and the "poor". The former ones get income both from working and from renting capital good; moreover, they are able to choose how many hours to work and how much to save. The latter ones neither own capital good nor save: they work a fixed number of hours and consume the whole income.

It should be noted that in our bargaining model the only way to save consists on buying newly produced capital goods which will be employed for production in the next Hicksian "week"; therefore, the act of saving is equivalent to moving a certain amount of income one single "week" in the future. Furthermore, the type of agent heterogeneity included in the model shows a qualitative nature, and it is not based simply on the value range of the parameters. In this sense, we can talk about a truly structural heterogeneity, rooted on distinct social roles of the agents.

---

[12] The computer application has been coded in C++ by the author, using Microsoft Visual C++ 6.0.

As far as the sequential behavior of the model is concerned, it can be sketched in the following way. At the beginning of each "week" the amount of productive resources and the prices of the three commodities are inherited from the previous time interval; the "rich", the "poor" and the two representative producers select their optimal behavior and put forward their ex-ante supplies and demands. Market balance is obtained by the usage of quantitative rationing, following the disequilibrium rule of the "short side", i.e. for each commodity the ex-post solution quantity coincides with the minimum between ex-ante demand and supply magnitudes. At the end of the "week" the newly produced commodities are available to the agents and after that the new prices are calculated, using the standard Walrasian adjustment rule based on the level of the ex-ante excess demand functions. Then a new Hicksian "week" takes place, following the same course of action.

Let's now proceed with some formal specifications. Agent (a), representing the "rich", optimizes his behavior by solving the following system:

```
System (1)
```

$$\text{MAX}\ \ U = C * D\_a\_c^{alfa\_one} * D\_nk^{alfa\_two} * Free\_Time^{alfa\_three}$$

$$\text{SUB}\ \ O\_ok * p\_ok + O\_a\_l * p\_w = D\_a\_c * p\_c + D\_nk * p\_nk$$

where:

`C, alfa_one, alfa_two, alfa_three` are the Cobb-Douglas utility function coefficients. Constant returns to scale imply: `alfa_one + alfa_two + alfa_three = 1`. The symbol `U` refers to the level of utility; `D_a_c, D_nk,`

`Free_Time` are the demands respectively of consumer good, new capital and free time; `O_ok`, `O_a_l` are the supplies respectively of old capital and labour; `p_ok`, `p_w`, `p_c`, `p_nk` are the prices of the corresponding magnitudes.

Agent (b), representing the "poor", doesn't have any choice but complying with his own balance constrain:

```
D_b_c = (O_b_l * p_w) / p_c
```

(2)

```
O_b_l = omega
```

Where `D_b_c` , `O_b_l` are respectively the demand of consumer good and the supply of labour; `omega` is the fixed amount of labour which meets the requirements of the standard employment contract.

As it's already been specified, both producers' technologies are represented by Cobb-Douglas production functions showing constant returns to scale, thus complying with the following formula:

(3)   `Product = B * Capital`$^{beta\_one}$` * Labour`$^{beta\_two}$

where the meaning of the symbols is intuitive. Following the standard Walrasian procedure, for each market aggregate demand and supply quantities can be calculated. At the end of the "week", the "short side" rule will determine the quantity to be actually produced by the following principle:

(4)   `Produced Quantity = minimum{ex-ante Demand, ex-ante Supply}`

Price adjustments at the end of the "week" will be carried out using the following procedure:

```
(5)   price(t + 1) = price(t) + {price(t) * 2 *atan[Demand(t) –
      Supply(t)]} * varmax
```

where `t` is the time specification of the "week" under examination; `atan[.]` is the arc tangent function and `varmax`(0 < varmax < 1) is a parameter setting the adjustment speed. It should be noted that expression (5) describes a price adjustment process which is a rather faithful formalization of the traditional Walrasian one, with the additional assumption that prices would always be strictly positive.

Using the above defined system, two different type of simulations have been carried out. In the first one, only the "rich" are present, while in the second one both the "rich" and the "poor" coexist in the same economic framework. Hence, the difference between the first and the second simulation is basically a matter of agent heterogeneity: in the first case there is only one kind of agent who is able to fully determine his own personal policy. In the second case, another type of agent is added, characterized by been unable to maximize his own utility: the "poor" must take life "as it comes". Both simulations have been tested with a large set of initially given conditions. The outcome is striking: in the first circumstance, after a finite number of "weeks" the economy collapses, since the "rich", while becoming richer, tent to offer less and less labour. At the end, they offer none of it. The latter simulation shows a completely different behavior. "Week" after "week", the "rich" still offer less labour, till the point when they offer none; but the "poor" will continue to work at the standard level and this amount of labour, combined with an ever increasing quantity of capital, is able to insure an everlasting growth of the economy. It should be noted that in the second case, while the "rich" obviously get richer and richer, the "poor" too enjoy the same treatment: they get less poor, while maintaining the same sociological role. In fact, the ratio between the wage and the price of the consumer good tends to increase indefinitely.

The results of the two simulations have been resumed in four graphs. In Graph (1), relative to the first simulation, there have been shown the quantities of capital and labour employed in the economy. After a finite number of "weeks", the latter becomes null, thus ending any economic activity.

The second simulation is illustrated by the usage of three graphs – Graph (2), (3) and (4). The outcomes support the above mentioned interpretation. Graph (2) shows that, even if from a given "week" on only the "poor" offer labour, the level of produced capital augments with an increasing rate of growth. Consequently, Graph (3) confirms that the global consumption level follows a similar growth pattern. Finally, Graph (4) guarantees that also the "poor" get more affluent: the ratio between the wage level and the price of the consumer good is an ever increasing digit.

It should be noted that a third simulation, relative to a contest were only the "poor" exist, has not be carried out, since obviously such an economy is due to collapse, because in such a contest capital cannot be reproduced. In conclusion, the assumption of agent heterogeneity is the only one which can guarantee unlimited growth. Both cases of agent homogeneity will eventually imply the collapse of the economy under consideration. Finally, a "steady state" was never obtained, thought of course we cannot rule out the possibility that such an event might occur if the initially given parameters assume the "right" values.

**Simulation 1 - Graph 1**

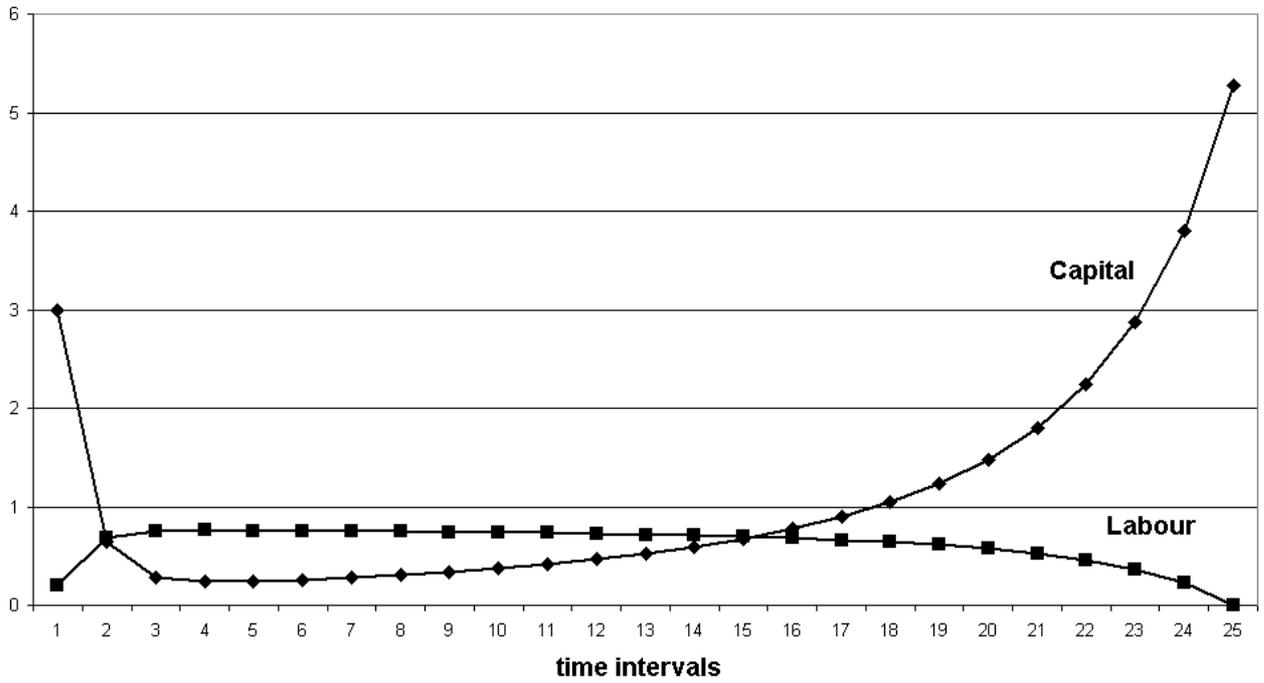

**Simulation 2 - Graph 2**

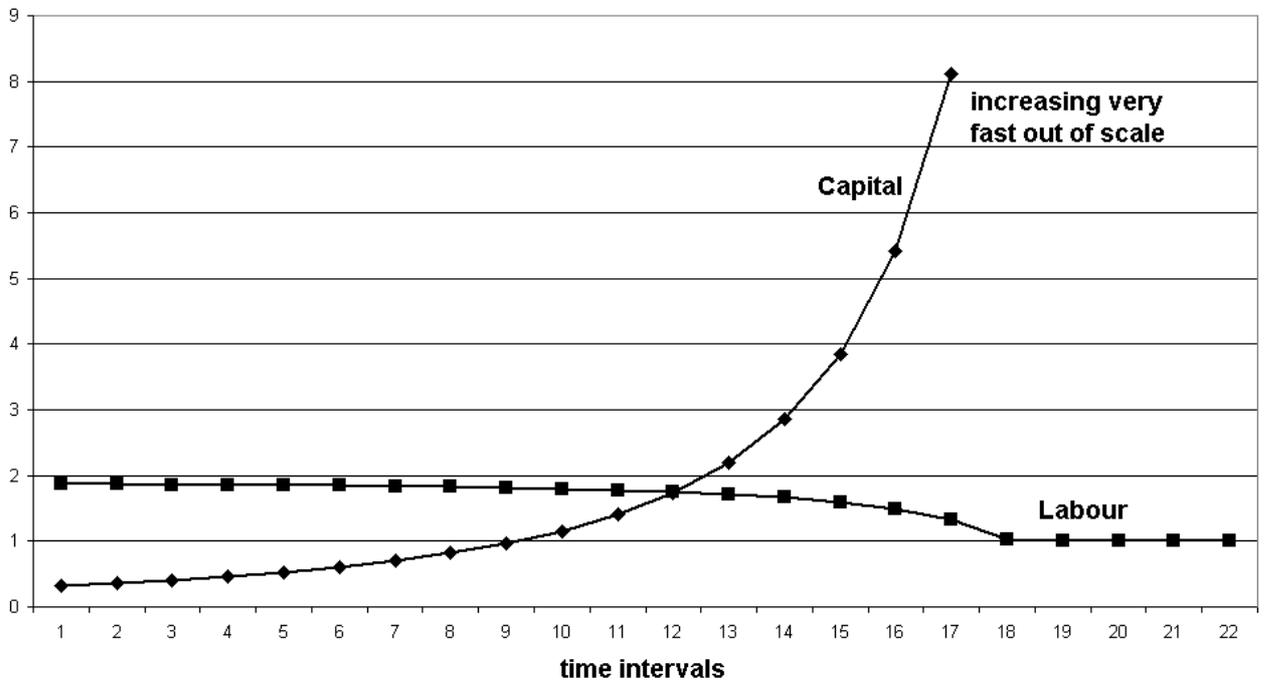

**Simulation 2 - Graph 3**

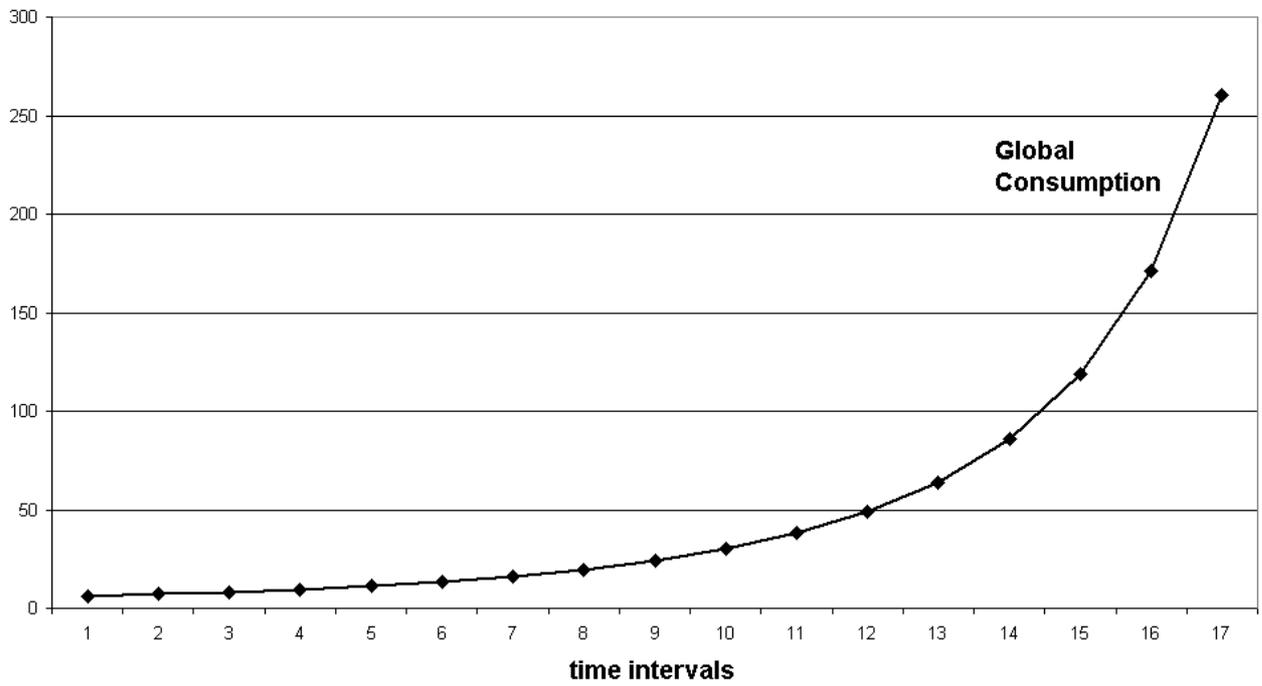

**Global Consumption**

time intervals

**Simulation 2 - Graph 4**

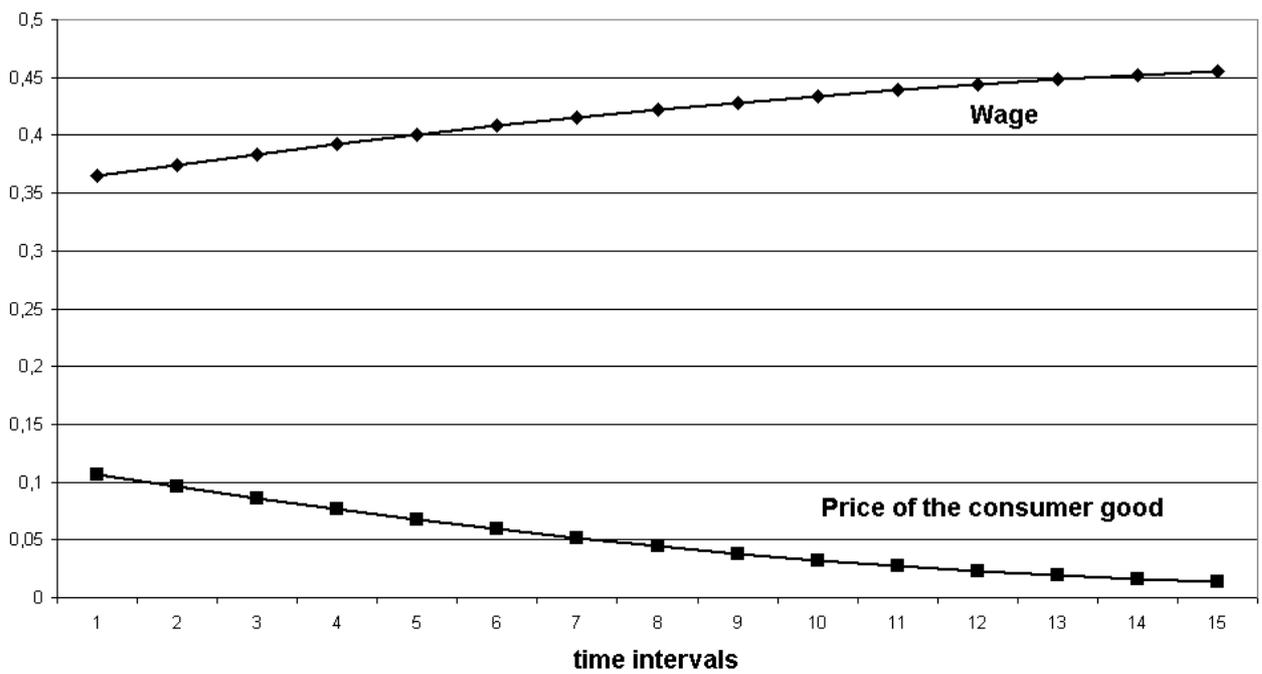

**Wage**

**Price of the consumer good**

time intervals

## 4. Does the general equilibrium theory belong to the gravitational realm?

Those who are familiar with the general equilibrium literature[13], are well aware that the problem of uniqueness and stability of the solution for the general equilibrium model has always be considered of crucial importance in order to assert the validity of that same theory. It is a well known fact that recently every hope to reach such a result has been relinquished, at least if we seek a sufficient condition which is based on microeconomic properties[14]. On the other hand, the formal consistency of the Arrow-Debreu approach has been proven beyond any doubt. It seems that such a line of research has reached its natural limits, as far as both positive and negative results are concerned. Depending on the point of view, such a situation can be considered more of less satisfactory, even if there is no doubt that what has been obtained is somehow less than what was expected when this line of economic thought started.

Still, reflecting upon what has been illustrated in the previous paragraphs, the following considerations can be put forward.

1. Uniqueness of the solution is basically a technical detail, more than a theoretical property. As Ludwig Wittgenstein has shown us, "the world is the totality of facts, not of things"[15]. Therefore, every single fact is unique by definition, in the sense that it can only occur or not. If two facts happen at the same time, that does not mean that each one isn't unique but simply that the environment we are considering is more complicated than we thought. Thus, it could be said that if a model doesn't imply the existence of a unique solution, when such a feature is sought after, then it is simply a bad model. Of course, within the general equilibrium theory the point is rather different: there has been the attempt to prove uniqueness, as well as stability, of the solution starting exclusively from the micro-foundation of the equilibrium system. At least in the framework of the Arrow-Debreu approach, such a task cannot be achieved. The example in the previous paragraph shows that if we are not worried with micro-foundation problems, uniqueness of the solution can be easily obtained even within rather simple and traditional models.

---

[13] We assume that the reader is familiar with the state of art of the general equilibrium theory, since the literature is so vast that it cannot be summarized here.

[14] Such is the consequence of the Debreu-Sonnenschein-Mantel theorem.

[15] Wittgenstein L., *Tractatus Logico-Philosophicus*, 1.1.

2. The problem of stability of the solution makes sense only in a gravitational contest, while in an evolutionary one it is completely meaningless. In the second approach, prices, as well as any other variable, change through time. Moreover, stationary or asymptotically converging patterns, as well as "steady states", can occur only by chance. Such circumstances, which within a gravitational approach are the only one considered worth studying, from the evolutionary point of view are just meaningless flukes. A recent contribution[16] has shown that within the Arrow-Debreu model instability, as well as the appearance of solutions with "undesirable" features, can be triggered by the presence of capitalization processes. Therefore, by a cause that is independent from the micro-foundation issue. As a consequence of such a finding, we are entitle to suspect that the problem does not lay with the choice to seek a micro-foundation for the already specified sufficient condition. On the contrary, we might hold the belief that the formal structure of the Arrow-Debreu model is unfit to provide such a goal. Once again, it could be a case of bad modeling.

Let's now concentrate our attention on the following question: can the Arrow-Debreu theory be considered a gravitational one? If we take into consideration the formalization which has been put forward by Gerard Debreu in *Theory of Value*, then the answer is no, since uniqueness and stability of solutions cannot be guaranteed by conditions which show an acceptable level of generality. Still, we know by experience that we can built specialized sub-models, i.e. models based on restrictive assumptions, in such a way to obtain gravitational environments. Therefore, we could conclude that the Arrow-Debreu approach should fall within the category of the weak gravitational environments. Such a case, which can be considered as a frontier between the gravitational and the evolutionary worlds, is too weak to fulfill the requirements needed by computable models. In this sense, the Arrow-Debreu paradigm can be viewed as an abstract framework useful to theoretically support operative instruments, rather than a model in itself. More a form of specialized scientific language rather than a theory in the proper sense of the world, since the way it is structured does not allow a direct interpretation of the world phenomena. In order to achieve such a power, the Arrow-Debreu framework must be adjusted to the actual portion of the economic word we intend to analyze. From this need it derives the huge amount of specialized equilibrium models which can be found in the literature[17].

---

[16] See Garegnani (2000).

---

[17] See Tucci (1997).

**5. Some conclusive observations**

The following considerations belong to three different fields of analysis. The first one concerns the comparison of the simulation results with the phenomena of the economic world. The second one is centered on the role played by the heterogeneity of the agents. Finally, the last one goes back to the methodological characterization of the evolutionary and gravitational environments.

Starting with the first issue, thought the above shown simulations are far too simplified to be used for analyzing world facts, we cannot escape the impression that the model points to a well known feature of the advanced economies: the structural shortage of labour in the lower section of the market. In fact, in our presentation the "rich", who enjoy a capital income and can compare the advantages of working with the pleasure of leisure time, will refuse jobs in the lower bracket, while of course the "poor" have no choice. In developed economies the population tents to became richer and richer, thus inducing a shortage of offer for the most disadvantageous jobs. Obviously, in order to be able to assert the significance of the above sketched procedure with respect of analyzing labour market dynamics, the structure of the simulations should be enriched with realistic details. At the present, we can only draw attention to an apparently illuminating behavior of the model.

Coming to the second issue, the relevance of agent heterogeneity with respect of the nature of the economic dynamics is strongly stressed. It should be noticed that the two classes of agent included in the model are truly heterogeneous in the qualitative sense of the term, given that they are not distinguished simply by the numerical value of the relative parameters. The role played by agent heterogeneity with respect of the outcome of the simulations is an utterly critical one: if only one type of agent is present the economy collapses, while if both are operating we will obtain an everlasting growth. As in the previous case, it would interesting to include in the model more detailed types of heterogeneity, in order to increase the level of realism of the analysis.

Finally, let's add some methodological consideration concerning the evolutionary and the gravitational approaches. The above shown simulations cannot be considered fully evolutionary from the Darwinian viewpoint, since a crucial feature of that theory, i.e. the process of mutation – selection, is missing. Still, the model is rather evolutionary in comparison with the standard gravitational literature. As it is well knows, the state of art of gravitational growth analysis implies a generalized usage of the "steady states" method and of approaches centered on asymptotic behaviors. Within a gravitational framework, only contexts showing such properties are taken into consideration, while every other circumstance is ignored, because is not considered suitable to lead to a solution which can be labeled as an "equilibrium". Now, it is obvious from the related graphs that the environment considered in the second simulation, while been economically meaningful, cannot be analyzed in terms of "steady states", since it implies an everlasting growth with a constantly increasing rate. Therefore, from the orthodox point of view the simulation should be considered of no relevance, while on the contrary it is obvious that such is not the case, since, as it has already been emphasized, there exist occurrences which seem to follow a pattern of behavior analogous to the simulated context. In conclusion, if we restrict the capability of our minds by assuming that only gravitational frameworks are acceptable, then we should not be surprised by our lack of understanding concerning whole classes of economic phenomena.

Finally, let's concentrate our attention on Poincarre's conception about the link between models and phenomena. If we agree that models are essentially tools to deal with events and consequently ought to be judged by the results and not by using *a priori* criteria, then we should give a fresh look to the great wealth of theoretical structures that the schools of economic thought provide us.

If we relinquish the belief that a single approach is "the right one", while every other one is "wrong", then we could come to agree that certain phenomena are better explained by a model coming from a specific research tradition, while other occurrences might be understood by the usage of a framework belonging to the rival school. What has been carried out in the present note is an example of such a statement. The evolutionary and the gravitational paradigms are rival approaches – if we go back to the Greeks, it's a matter of Heraclitus versus Parmenides. Still, in the present note it has been shown that both methodologies are valid in interpreting economic phenomena, each one giving its best in certain contexts, while been ineffective in others. A multilateral approach seems more a necessity than a choice when we come to deal with the complexity of the word economy.

# Bibliography[18]

---

[18] The following bibliography is more vast than the content of the present notes, since it refers to topics belonging to the research project quoted in the introduction.

*Post Scriptum*

# Evoluzione Gravitazione Economia

## Archetipi

Evoluzione... Nella sua forma estrema, una delle idee più terrificanti concepite dalla mente umana. Il vento del deserto trasforma la pietra in polvere. Il divenire sgretola le grandi concezioni, le visioni universali, i rassicuranti precetti tramandati dai nostri padri. In un mondo costantemente rinnovato, ogni evento costituisce una realtà che nasce e muore nell'ambito della propria singolarità, nell'assoluto della propria limitatezza. Cogli l'attimo fuggente! Ma come è possibile afferrare qualcosa la cui natura è fugacità? Se l'acqua che scorre nel fiume non è mai la stessa, sempre diverse sono le gocce che bagnano la riva. E che dire del vapore che si perde nell'aria? Le entità diventano evanescenti, soffuse di nulla. Esiste forse il fiore azzurro sbocciato questa mattina e già appassito? Ed il bocciolo che domani sarà una rosa e dopodomani un grumo di materia inerte? Quanto spazio c'è in un mattino? E quanto tempo in una moneta da cento lire? Quanti cavalli in Tessaglia? A corto di leggi, di classificazioni, di vocabolari, di linguaggi, la mente annega nel grande oceano primordiale, sperimentando le estasi di infiniti paradisi ed i tormenti di infiniti inferni. E' lo studioso che pensa più reale del contenuto dei suoi pensieri? O non è il soggetto stesso una categoria transeunte, rugiada su un filo d'erba, in attesa di evaporare al primo sole dell'alba? Nulla dunque esiste. Tutto scorre. L'universo è la cascata di faville di un fuoco d'artificio.

Gravitazione... Nella sua forma estrema, una delle idee più terrificanti concepite dalla mente umana. Al centro del mondo, esiste un punto che racchiude l'assoluto di tutte le verità. Da tale radice eterna fluiscono filamenti sottili, raggi di consistenza eterica che raggiungono i più remoti grumi di materia. L'universo è reale solamente in quanto è permeato dalle emanazioni del suo centro immutabile. L'unico problema che la mente umana possa risolvere, consiste nel trovare una adeguata formulazione linguistica del seme originario, il nome vero... Pochi segni in grado di racchiudere ciò che è eternamene. Ogni invenzione, ogni sforzo del pensiero, non può, nel migliore dei casi, che riportarci là dove già eravamo, di fronte al nostro punto di partenza. Ogni movimento è illusorio, ogni pensiero degno di questo nome consisterà esclusivamente in una catena di deduzioni, in grado di derivare verità parziali dall'unica verità globale, sempre la stessa. Il tempo è illusorio, poiché ogni istante è già ben definito da sempre e per sempre. Il pensiero è una foresta di cristallo, scintillante ed immutabile. E' forse libera la mente che pensa? Oppure non è anch'essa una sequenza di fotogrammi che si susseguono nell'unico ordine possibile, incatenati alle ferree leggi della causalità? Ogni cosa dunque è soggetta alla norma. Niente è libero. L'universo è un pesce rosso imprigionato in una sfera di ghiaccio.

**Economia**

Spesso, nel corso del tempo, il pensiero economico ha mostrato un certo spirito di emulazione verso l'apparato concettuale e gli strumenti matematici utilizzati dalle scienze naturali. Del resto, ciò non può destare particolare sorpresa, poiché nel mondo occidentale quest'ultimo settore di indagini, insieme con i correlati sviluppi tecnologici, ha rappresentato il motore del processo di crescita globale. Ispirandosi alla legge di Newton, generazioni di studiosi hanno cercato l'essenza ultima del valore economico, attraverso la formulazioni di teorie ciascuna delle quali aspirava al ruolo di verità di ultima analisi, valida sempre e ovunque. La gran mole di studi scaturita da tale approccio ha contribuito ad edificare le fondazioni del pensiero economico. Tuttavia, occorre notare che nessuna formulazione è stata in grado di avvicinarsi al livello di capacità predittiva che caratterizza la legge di gravitazione universale. Tale evenienza non può essere attribuita all'inadeguatezza dell'apparato matematico, il quale peraltro si è dimostrato notevolmente efficace nell'ambito delle discipline concernenti i fenomeni naturali. La scarsità di abilità previsionali, che affligge i filoni di ricerca in questione, va attribuita a cause assai più profonde e strutturali. A seguito di riflessioni concernenti le radici della problematica, alcuni studiosi hanno ritenuto opportuno far riferimento a Darwin ed al suo approccio, dialetticamente antitetico a quello gravitazionale. Di qui il fiorire di sviluppi teorici, rimarchevoli per raffinatezza dei costrutti ma, in qualche modo, viziati dal sospetto di una generale vaghezza delle fondazioni. Tale dubbio non è del tutto infondato, poiché è abbastanza evidente che una teoria di tipo evolutivo non possa basarsi interamente su proposizioni di natura deduttiva, come è il caso degli approcci gravitazionali. La mancanza del centro di gravità implica necessariamente un maggior ricorso a costanti esogene localmente determinate e dunque, alla fin fine, la strutturazione dei modelli risulterà meno logicamente concatenata di quanto non accada per teorizzazioni di altra natura. Se ci si spinge troppo lontano sulla strada della costruzione di teorie valide solo localmente, si rischia di costruire una scienza che coincida con l'immagine del mondo percepita dalla profanità. Ovvero, una entità sostanzialmente inutile. Viceversa, c'è da notare che, ove l'oggetto dell'analisi sia definito con sufficiente chiarezza, la teoria economica ha sviluppato metodologie in grado di fornire previsioni con un alto grado di affidabilità, così come di regolare il corso degli eventi mediante operazioni di *fine tuning*. In conclusione, lo stato dell'arte sembra indicare una soluzione di compromesso. Accantonata l'aspirazione agli assoluti, evitata la tentazione ridurre il mondo ad una miriade di frammenti privi di leggibilità, nell'ambito di scenari sufficientemente vasti, e tuttavia ben delimitabili, la scienza economica è in grado di esprimersi al meglio. Tale è lo stato dell'arte. Qualcuno è in grado di andare oltre?